\pdfoutput=1
\documentclass[twocolumn,showpacs,preprintnumbers,amsmath,amssymb, prl]{revtex4}

\usepackage{dcolumn}
\usepackage{bm}
\usepackage{hyperref}
\usepackage{amsmath}
\usepackage{amsfonts}
\usepackage{graphics}
\usepackage{psfrag}
\usepackage{graphicx}
\usepackage{epsfig}
\usepackage{rotating}
 \usepackage[latin1]{inputenc}
 \usepackage{color}
 \usepackage{rotating}
 \usepackage{threeparttable}  
 \usepackage{multirow}        
 \usepackage{color}
\usepackage{lscape}

\definecolor{DarkGreen}{rgb}{0,0.5,0}
\definecolor{Grey}{rgb}{0.5,0.5,0.5}
\definecolor{DarkYellow}{rgb}{1,0.7,0}
\definecolor{Violet}{rgb}{0.6,0.0,0.7}
\definecolor{Brown}{rgb}{0.5,0.3,0}

\begin{document}


\title{Measurement and Application of Entropy Production Rate\\ in Human Subject Social Interaction Systems}
\author{Bin Xu}
\author{Zhijian Wang\footnote{Corresponding author. This manuscript is modified from its early version in ~http://ssrn.com/abstract=1878063.}}

\affiliation{Experimental Social Science Laboratory, Zhejiang University, Hangzhou, 310058, China}



\date{\today}
\begin{abstract}
This paper illustrates the measurement and the applications of the observable, entropy production rate (EPR), in human subject social interaction systems. To this end, we show (1) how to test the minimax randomization model with experimental 2$\times$2 games data and with the Wimbledon Tennis data; (2) how to identify the Edgeworth price cycle in experimental market data; and (3) the relationship within EPR and motion in data. As a result, in human subject social interaction systems, EPR can be measured practically and can be employed to test models and to search for facts efficiently.
\end{abstract}



\pacs{87.23.Cc  
89.65.-s 
01.50.My 
02.50.Le 
}
\maketitle
\section{Introduction}

Laboratory experiment in human  subject social interaction systems (HSSIS) has becoming major tool for fundamentally social science~\cite{Falk2009,Plott2008}, in which developing signature observable is meaningful.

Entropy and entropy production rate (EPR), a twin observable, correspond to the diversity and the activity of a system~\cite{Frey2010,Schmittmann2007}, respectively. In game theory for HSSIS, entropy has been noticed theoretically \cite{Entropy2004theo} and experimentally~\cite{Yan2011,Cason2009}. Denoting the density of state (DOS) as $P_i$ for state $i$ ($i\in X$ and $X$$:=\{1,2,...,r\}$ is the full social strategy set), the entropy $S$ is as~\cite{Yan2011},
\begin{equation}
S=-\sum_i P_i \log_{r} P_i.
\label{eq:entropydefine}
\end{equation}
In games, entropy can identify the distribution in $X$ and the diversity of  HSSIS~\cite{Yan2011,Cason2009}.
EPR serves as a central observable for the activity of many natural systems~\cite{Frey2010}, however, in HSSIS, EPR has never been reported empirically; Developing EPR as an observable in HSSIS is the main aim of this letter.

\begin{figure}
\centering
\includegraphics[angle=0,width=6cm]{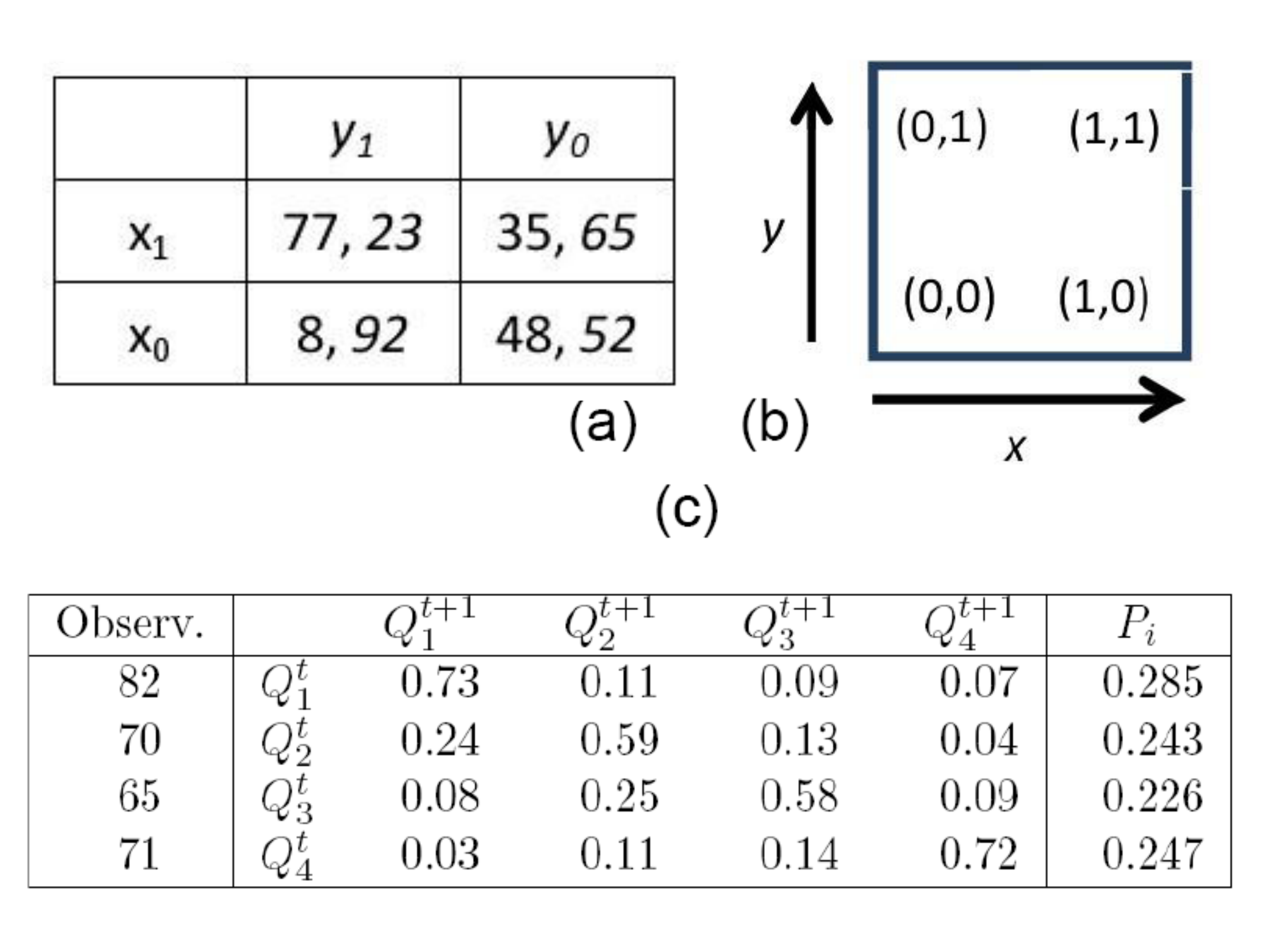}
\caption{(a, b) a payoff matrix and the strategy space $X$ (2$\times$2 game, Trt. 31 in Table~\ref{tab:DataList}); (c) a Markovian with $\omega_{ij}$ and $P_i$ (from top-left subtable in
Table 1 in ref.~\cite{cason2005dynamics}, also Trt. 48 in Table~\ref{tab:DataList}); in (c), the entropy is 0.9952 and the EPR is 0.0481.}
\label{fig:Payoff4States}
\end{figure}

As the discrete Markov chains can be obtained~\cite{cason2005dynamics,Buchheit2010} in HSSIS,  the metric for EPR could borrow  from physics~\cite{Gaspard2004,Maes2003,Frey2010}.
 For a system with small number of states and lasting time long enough, the stationary state approximation can be considered~\cite{selten2008}; the associated mean \emph{entropy production rate} $\dot{S}$ (EPR) is as~\cite{Gaspard2004,Maes2011,Pleimling2011,Frey2010}
\begin{equation}
\dot{S} = \frac{1}{2}\sum_{i,j} \left[ P_i \omega_{ij} - P_j \omega _{ji}\right] \log_r \left [\frac{P_i \omega_{ij} }{ P_j \omega_{ji}}\right ]\label{eq:defineEntropyProduction},
\end{equation}
%
%
%
in which $P_i$ is the DOS and $\omega_{ij}$ is the transition probability from state $i$ to $j$.
Fig.~\ref{fig:Payoff4States}(c) is an empirical example for $P_i$, $\omega_{ij}$ and a Markovian.


\begin{table}[htbp]
\caption{Empirical Data from HSSIS}
\label{tab:DataList}
\centering
\begin{tabular}{|c|c|c|r|c|}
  \hline
	  ~~ Game~~ &~ Trt.\footnote{Trt.: Treatment, S.: Session, Rec./Trt.: Records in per treatment, 2$\times$2: the 2$\times$2 games, WT: Wimbledon Tennis~\cite{Hsu2007}, EC: Edgeworth Cycle~\cite{cason2005dynamics}, HS: a laboratory 2$\times$2 game~\cite{Jason2003} and MP4: our laboratory experiment with 12 sessions 300 rounds fixed paired 2-person matching pennies game with payoff matrix as $[5,0;0,5;0,5;5,0]$ as~\cite{XuWang2011}. Notices that, WT data is not an experimental economics (EE) data but records of Tennis game.}~	 & ~States~	&~Rec./Trt.~& ~~~Ref.~~~	\\	 \hline
 	   2$\times$2	&31-40 	&4	&4500~~	&\cite{RothErev2007}	\\	
	   2$\times$2	&41	&  4& $\sim$3600~~	& MP4	\\	
	  WT(2$\times$2)	        &45-47 	&4& $\sim$2000~~	&\cite{Hsu2007}	\\	
	  EC	        &48-55	 	&4&$\sim$200~~	&\cite{cason2005dynamics}	\\
	  HS(2$\times$2)	        &56-57	 	&4&  2600~~	&\cite{Jason2003}	\\	\hline
\end{tabular}
\begin{flushleft}
\end{flushleft}
\end{table}

The data is collected from the published 24 experimental economics treatments (EET) listed in Table~\ref{tab:DataList}. All of the 24 EET, the state number ($r$) is 4. Each of the EET is of an unique mixed strategy Nash equilibrium and of long rounds satisfying the stationary approximation~\cite{selten2008}.  These are practically the simplest systems for EPR~\footnote{For a 2-state system, there is no possible to obtain EPR. The smallest system is of 3-state, e.g., Rock-Paper-Sessior game~\cite{Frey2010}. However, from laboratory or field experimental economics date in published literatures, no enough 3-state data with widely parameters and long rounds experimental sessions could be collected till now.}. For more details, see Appendix.

To demonstrate the EPR's applications, the paper is as follows, (1) testing an economics model with EPR; (2) detecting an economics phenomena with EPR; and (3) verifying a relationship within two dynamical variables with EPR; Then, summary last.

\section{EPR and Radomization in Minimax}

The von Neumann's Minimax model (vNM) \emph{'represents game theory in its most elegant form: simple but with stark predictions'}~\cite{Levitt2010}. In vNM, playing mixed strategy game \emph{'against an at least moderately intelligent opponent'}, should be
 '\emph{playing irregularly  heads  and  tails in successive
games}' (the randomization prediction), except that the possibility of a strategy used
is constrained. The randomization prediction of vNM is tested with EPR.

\begin{figure}
\centering
\includegraphics[angle=0,width=6cm]{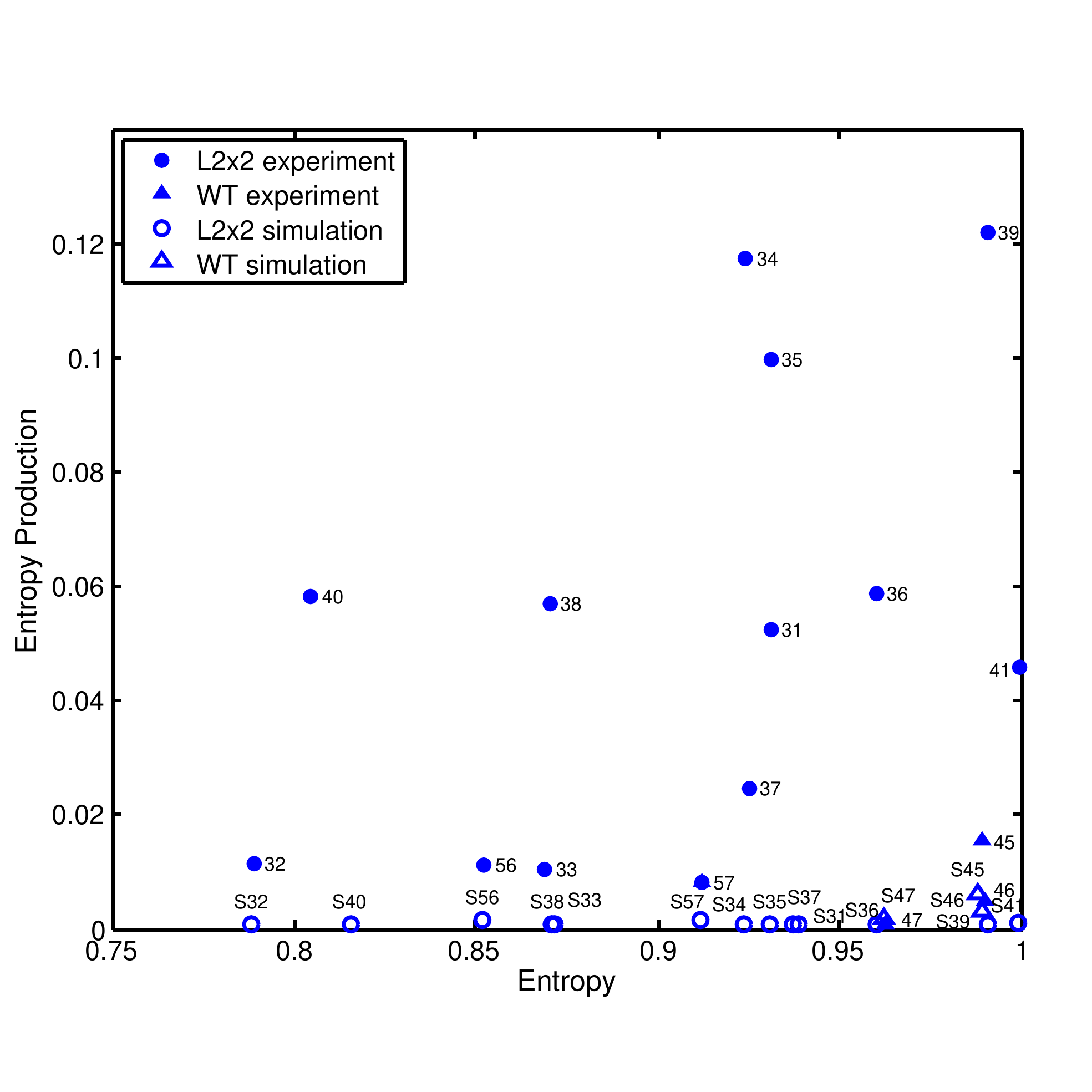}
\caption{Obtained entropy  (in horizon) and and EPR (in vertical) in the 2$\times$2 games. Experimental results of the Trt. ID in Table~\ref{tab:DataList} labels in solid.  The result of mean entropy and EPR from the $10^4$ times vNMs are in hole and labeled start with 'S' and the Trt. ID.}
\label{fig:EntropyProductionRoth14}
\end{figure}

 In Fig.~\ref{fig:EntropyProductionRoth14}, the entropy and EPR from the 16 EET of the 2$\times$2 games (Trt.  31-41, 45-47 and 56-57) are shown (in $\bullet$ or $\blacktriangle$).

The vNM randomization predictions can be realized directly with simple Monte Carlo simulation (MCs)~\cite{Palacios2003}.
We conduct the MCs $10^4$ times for each of the 16 EET, respectively; For a MCs,  there are two constrains: First, holds the same sample size of sessions, rounds and agents as its related EET; Second, holds the same mean strategy possibility as its related EET~\footnote{Fig.1 (b) illustrates the state space $X$ of a two-person 2$\times$2 game and $X$=:$\{($0,0$), ($0,1$), ($1,0$) ,($1,1$)\}$. In a experimental treatment $k$, the mean observation $\bar{x}$=($p,q$)$_k$ in the 1$\times$1 square can be obtained~\cite{selten2008,RothErev2007}.
For MC artificially generated data for related treatment $k$,  the mean strategy possibility constraints is keeping the ($p,q$)$_k$ as the possibility to choose (Up, Left)-strategy for the two agents respectively at each of the simulation round. The random Monte Carlo (uniform distribution in $[0,1]$) seeds are generated with Matlab 2010a.}.
 As results of the vNM randomization prediction, the entropy and EPR~\footnote{For more information on the entropy and EPR from MCs for each of the treatments see Appendix.} are shown (in $\circ$ or $\vartriangle$) in Fig.~\ref{fig:EntropyProductionRoth14}.

Comparing entropy and EPR  values from the vNM and from the HSSIS respectively, we have (1)
 the vNM can not be rejected with the entropy comparison~\footnote{At $p<0.05$ standard, with MC $10^4$ samples for each of the treatments, 8 of which the entropy larger than that from HSSIS, 6 smaller and 3 is indistinguishable;  $t$-$test$.}; But, (2) the EPR  values from HSSIS is significant larger than that from vNM ($p$$<$0.001,  16 samples, $t-test$~\footnote{For each of the treatments, except the Trt. 47 of which is the male players of the Wembelden Tennis, the EPR from HSSIS is larger significant.}). With EPR, the randomization prediction of vNM have to be rejected.


It is an example of testing a behavior model with EPR.

%
%
%

\section{EPR and  Edgeworth Cycle}

Many years ago, Edgeworth predicted persistent price cycles phenomena in a competitive situation where the only equilibrium is in mixed strategies~\cite{benaim2009learning,Maskin1988}. It is very puzzling because
it seemingly contradicts the \emph{law of one price} of elementary microeconomics~\cite{Lahkar2007}. The cycles, e.g., in retail gasoline markets~\cite{Noel2007}, have been obtained in real economies; meanwhile, the welfare effects of the cycles have been found~\cite{Noel2011}. Usually, detecting the cycle is uneasy~\cite{benaim2009learning,cason2005dynamics}.


In a stationary state, if the EPR systematically deviates from zero, there must be balanced cycle fluxes, and this is a simple consequence of Kirchhoff's Law~\cite{Gaspard2004,qian2005}. Meanwhile, in a finite dataset, as Trt. 48-55, the obtained
EPR can be  the bias from finite sample (e.g.,~\cite{BiasEntropy2004}). To correct the bias, we use the EPR from repeated MCs as base line zero (denoted as $B^{0}$). So the existence of a cycle in a EET can be simplified to a sharp testable hypothesis (H0):  for the $i$-th EET, the empirical $EPR_i$ equals to the $B^{0}_{i}$.



\begin{figure}
\centering
\includegraphics[angle=0,width=6cm]{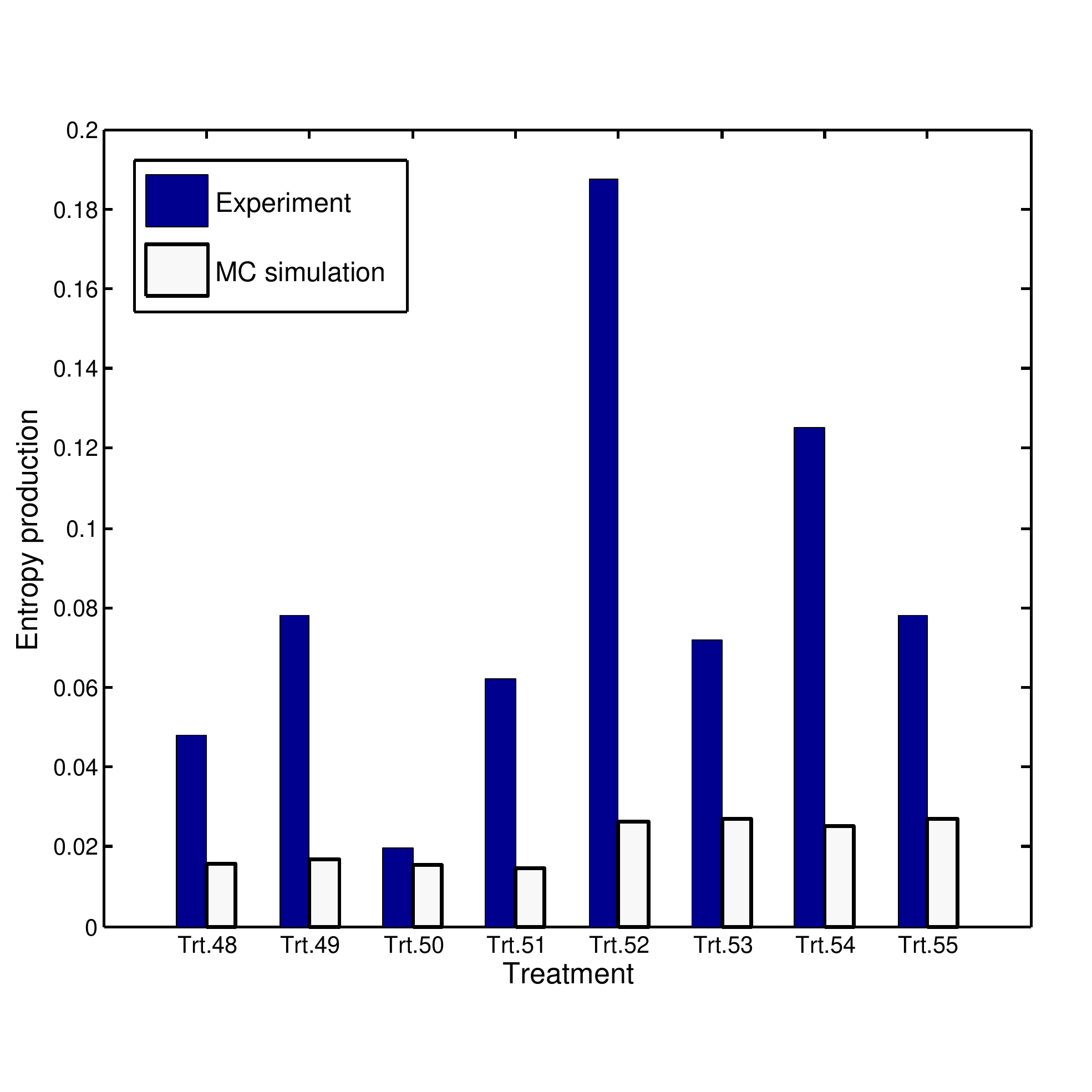}
\caption{EPR in the price dispersion market game from the data of the
laboratory experiments (solid bar) and Monte Carlo simulations (blank bar, mean value of $B^0_i$, $10^4$ times repeated, see text), respectively.}
\label{fig:EPMeanmiCasonPriceCycle}
\end{figure}

In Fig.~\ref{fig:EPMeanmiCasonPriceCycle}, each of the empirical EPR$_i$ of the 8 EET (Trt. 48-55) is shown in solid bar and
the $B^{0}_{i}$ is in blank bar.
Each $B^{0}_{i}$ comes from the MCs with the constraint of the price distribution from the Markovian and the number of rounds as its related EET. As a result, in 7 EET (except Trt. 50), the empirical EPR$_i$ is larger significant than its $B^{0}_{i}$ ($p<0.001$)~\footnote{One-sample $t$-$test$, empirical EPR$_i$ compares with the discrete $10^4$ samples of $B^{0}_{i}$ from MCs.}. So, the existence of Edgeworth cycle in each of the 7 EET can be supported efficiently~\footnote{In~\cite{cason2005dynamics,benaim2009learning}, to identify the existence of the cycle needs to pool all the 8 EET data together; and in EPR metric, the existence can be supported in each of the 7 EET.}.

It is an example of detecting an economic phenomena with EPR.

%
%
%

\section{EPR and Motion}



\begin{figure}
\centering
\includegraphics[angle=0,width=6cm]{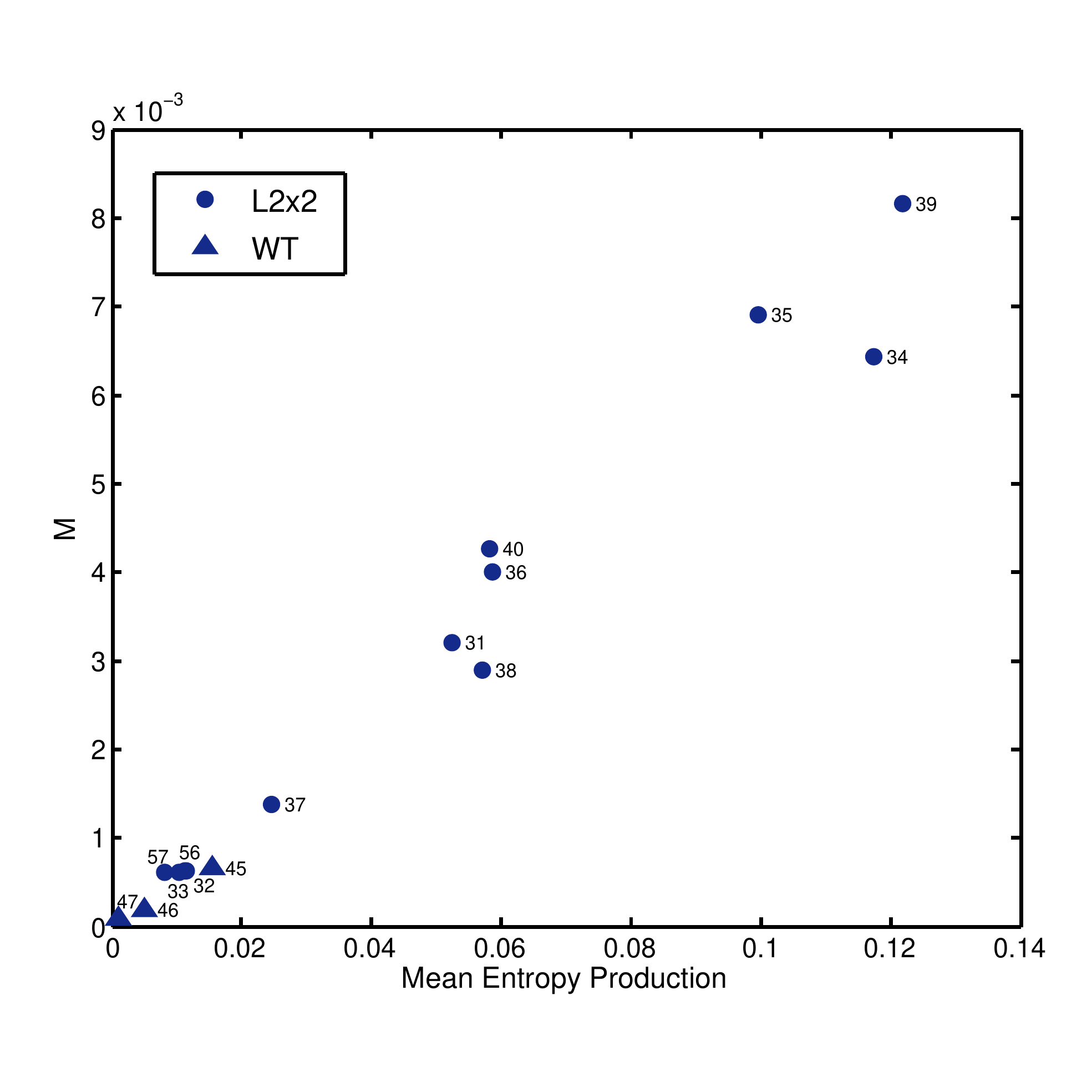}
\caption{Empirical relationship of the motion (M) and EPR in the EET of the 2$\times$2 games. Labels relate to the Trt. ID in Table~\ref{tab:DataList}}
\label{fig:MotionProduction}
\end{figure}

In evolutionary game theory, an evolutionary dynamic equation describes the velocity of evolution in space $X$~\cite{TraulsenPRE2009, Weibull1997,Bowles2004,Sigmund2010}.
Experimental economics is a test bed for evolutionary game theory~\cite{Crawford1991,Binmore1993,Samuelson2002, Huyck1995,Binmore2001, HuyckSamuelson2001, cason2005dynamics, Friedman2008}, and recently the velocity patterns in strategy space are observed~\cite{XuWang2011ICCS,XuWang2011}.
As both of the observables, the velocity in Eq.(\ref{eq:velocityInMM}) and the EPR  in Eq.(\ref{eq:defineEntropyProduction}), describe the time reversal asymmetry processes of experimental social dynamics
~\cite{Schmittmann2007,XuWang2011ICCS,Baiesi2009,Castellano2009}, the relationship within them is natural concerned.

Fig.~\ref{fig:MotionProduction} is the scattering of the motion ($M$) and EPR of the 16 EET (Trt. 31-41, 45-47 and 56-57, the 2$\times$2 games). Here,
the motion ($M$) is defined as,
\begin{equation}\label{eq:MDM}
    M=\frac{1}{2} \sum_{i\alpha} P_{i}v_{i\alpha}^2,
\end{equation}
 in which, $P_{i}$ is the DOS of $i$ ($i \in \mathbf{S}$) and $\alpha$ denotes the two dimensions of movements in a two-population 2$\times$2 games as in Fig.~\ref{fig:Payoff4States}(b). The velocity~\cite{XuWang2011ICCS,XuWang2011}, $v_{i\alpha}$, in Markovian format is
  \begin{equation}\label{eq:velocityInMM}
   v_{i\alpha}=\sum_{j}\left[ (P_j \omega_{ji} - P_i \omega _{ij})(x_{i\alpha} - x_{j\alpha})\right],
\end{equation}
in which $x_{i\alpha}$ is the vector of the $\alpha$ companion of the $i$ state in Euclidean $X$ ~\cite{XuWang2011ICCS,XuWang2011}.
Denoting the EPR  as $P$, the simple linear fit (OLE) results: $M$=($0.064\pm$0.003)$P$ + $0.000$ and $R^2 = 0.97$ for the 16 EET.
%
%

As an empirical finding, the motion ($M$) is positive and linear dependent on the EPR  in the data.

\section{Summary and outlook}

There is no reason that the approach to predicting the behavior of
physical systems is not appropriate when the physical system in question is some human
beings playing a game~\cite{Wolpert2010}. EPR is one of the key signatures of non-equilibrium steady (stationary) states~\cite{Schmittmann2007} and we link EPR to the stationary state~\cite{selten2008} of HSSIS.
The potential advantage of the EPR observable should be:

(1) \emph{At speaking to theorists}~\cite{Roth2010}: The EPR could serve as an independent variable to test behavior models. For example, testing the randomization prediction of vNM is hard ~\cite{O'Neill1987,Rosenthal1990,Binmore2001,Walker2001,Jason2002,Levitt2002,Levitt2010}, as we have shown above, EPR is effective for the task. Whether models~\cite{RothComp2010,selten2008} can be verified with EPR is becoming a question.

(2) \emph{At searching for facts}~\cite{Roth2010}: (a) EPR could serve as a variable to detect the heterogeneous of behavior, e.g. in Wimbledon Tennis~\cite{Hsu2007}, the EPR in the juniors, females and males are significant different. (b) the  EPR could detect the dynamical pattern. With the time reversal symmetry (asymmetry) consideration, unifying themes for the phenomena like Edgeworth cycle, Shapley polygon~\cite{Shapley1964}, and the Scarf price dynamics on Walrasian general equilibrium~\cite{Plottcycle2004,Gintis2007} can be expected.
%

(3) \emph{At Bridging between physics and economics}: Since 1990s~\cite{Evans1993}, the physics near stationary (or steady, equilibrium) state is becoming fruitful~\cite{Maes2003,Gaspard2004,Schmittmann2007,Pleimling2011,Maes2011}. This paper is benefit from the physics. Meanwhile, the empirical results of EPR and the methods here could feedback to the developing physics, e.g., to verify the relations of the observable (and variables). In social dynamics~\cite{Castellano2009}, with dynamical observable likes EPR, excellent developments could be expected.
 

%
%
%
%
%
%
%


 In summary, firstly and empirically in HSSIS, this paper has illustrated function of the observable, EPR, on models testing and facts finding, which could benefit to the both, physics and economics.

Our outlook is, as in physics, EPR can be a signature observable in the human subject social interaction systems.

\textbf{Notes}: We thanks Ken Binmore and Al Roth for helpful discussion and the data providing. The programmes
for the Monte Carlo simulations and statistical analysis, the primary data set and the instructions of our laboratory experiment MP4 in Table~\ref{tab:DataList} are available from the authors website.


%

%

\bibliography{mainMar0729}
\end{document}